\documentclass[journal]{IEEEtran}
\usepackage{graphicx} 
\usepackage{cite} 
\usepackage{amsmath,amsfonts,mathrsfs}
\usepackage{amssymb}
\usepackage{amsthm}  
\usepackage{booktabs} 
\usepackage{bbm}
\usepackage{setspace}
\usepackage{soul} 
\usepackage{color, xcolor} 

\def\E{\mathrm{E}}
\def\R{\mathbb{R}}
\def\d{\mathrm{d}}
\def\H{\mathsf{H}}
\def\F{\mathsf{F}}
\def\W{\mathsf{W}}
\newcommand{\erf}[1]{{\rm{erf}}\left(#1\right)}

\newcommand{\G}[1]{{G}_{{\rm{#1}}}}
\newcommand{\La}[2]{\mathcal{L}_{#1}\left(#2\right)}
\newcommand{\figref}[1]{Fig. \ref{#1}}

\newcommand{\RNum}[1]{\uppercase\expandafter{\romannumeral #1\relax}}
\newtheorem{theorem}{Theorem}
\newtheorem{corollary}{Corollary}
\newtheorem{lemma}{Lemma}
\newtheorem{proposition}{Proposition}
\begin{document}

\title{Performance Analysis of Millimeter Wave Wireless Power Transfer With Imperfect Beam Alignment}
\author{Man~Wang, Chao~Zhang,   Xiaoming~Chen,~\IEEEmembership{Senior Member,~IEEE}, Suhua~Tang,~\IEEEmembership{Member,~IEEE}
\thanks{Copyright (c) 2015 IEEE. Personal use of this material is permitted. However, permission to use this material for any other purposes must be obtained from the IEEE by sending a request to pubs-permissions@ieee.org.}
\thanks{M. Wang, C. Zhang and X. Chen are with School of Information and Communications Engineering, Xi'an Jiaotong University, Xi'an, Shaanxi 710049, China. (e-mail: manwang@stu.xjtu.edu.cn, chaozhang@mail.xjtu.edu.cn, xiaoming.chen@mail.xjtu.edu.cn)}
\thanks{S. Tang is with Department of Computer and Network Engineering, The University of  Electro-Communications,
	1-5-1 Chofugaoka, Chofu, Tokyo 182-8585, Japan. (e-mail:shtang@uec.ac.jp)}
\thanks{The work is supported in part by National Key Research and Development Program of China (No.2020YFB1807000), and the Key Research and Development Program of Shaanxi Province (No. 2020GY-023).}
}

\maketitle

\begin{abstract}
In this paper, the impact of imperfect beam alignment (IBA) on millimeter wave (mmWave) wireless power transfer (WPT) is investigated. We consider a mmWave WPT network, where the location of the energy transmitters follows a Poisson point process. Instead of the mostly used flat-top antenna model,  we adopt the Gaussian antenna model suggested by the 3rd Generation Partnership Project (3GPP) for better accuracy.   Two beam alignment error (BAE) models, i.e.,  truncated  Gaussian and uniform models, are introduced to represent different BAE sources.  We derive the probability density functions (PDFs) of the  cascaded antenna gain with both  BAE models and then  provide the approximated PDFs for tractability. With the help of Fox's H function, the analytic expression for the energy coverage probability with nonlinear energy harvesting model is derived. Besides, we deduce a closed-form expression of the average harvested radio frequency (RF) energy. 
The simulation results verify our theoretical results and demonstrate the performance degradation incurred by BAE. It also shows that in the imperfect beam alignment scenario, the Gaussian antenna model can accurately represent the performance of mmWave WPT networks with actual beam pattern, while  the flat-top antenna model cannot always provide accurate performance evaluation. 
\end{abstract}


\IEEEpeerreviewmaketitle

\section{Introduction}
\IEEEPARstart{I}{n} the foreseeable future, there would be huge amounts of low-power devices in wireless networks to perform information forwarding, data  collecting, situation sensing  \cite{BC_SWIPT}. These devices could be the nodes in Internet of Things (IoT), wireless sensor networks (WSNs) or Device-to-Device (D2D) systems. These low-power devices are usually powered by batteries  and  deployed in a broad  area \cite{Renzo_SL}. In order to prolong the network lifetime and maintain the sustainability of these nodes, far-field wireless power transfer (WPT) via radio frequency (RF) has been considered as a promising technology to energize massive battery-powered devices. The reason is that RF-WPT can provide a flexible and long-distance charging service, while being compatible with the existing wireless information networks \cite{Zeng_WPT, lu2017_survey}. At present, RF-WPT has been successfully applied in various wireless networks \cite{BC_SWIPT, Renzo_SL, Zeng_WPT, lu2017_survey}. 

For addressing the demand of ever-increasing data transmission rates,  millimeter wave (mmWave) frequencies are leveraged to fulfill the multi-Gigabit information transmission requirement  \cite{5Gchannel_model}. Due to the small wavelength and blockage sensitivity of mmWave signals, strong directional antenna and  small cell structure are suggested to enhance the energy and spectrum efficiency of mmWave communications \cite{blockage_model}.  As WPT also suffers from the severe power propagation loss and needs to avoid interference to existing wireless information networks, mmWave also benefits WPT.  It has been proven by \cite{mmWave_EH, howmany} that mmWave WPT outperforms WPT with lower frequencies.  
In \cite{Lens},  lens array based mmWave WPT was proposed to charge multiple energy receivers.  In \cite{DLL}, simultaneous wireless information and power transfer (SWIPT) was applied in the hybrid precoding mmWave system. Besides, the effect of rain attenuation on mmWave WPT was investigated by  \cite{rain}.  All that literature shows mmWave WPT is feasible.

To evaluate the system-level performance of mmWave WPT, stochastic geometry has been utilized to capture the effects of propagation loss and blockage, which dominate the received signal power  \cite{gamma_approx,Renzo}. 
The energy coverage probability and average harvested energy of mmWave WPT were studied by \cite{mmWave_EH}, where the location of  energy transmitters follows the  Poisson point process (PPP).  Furthermore, SWIPT was also introduced into stochastic mmWave networks \cite{howmany, Renzo_SWIPT}.  Both works verified that mmWave could improve the performance of SWIPT compared to lower frequency solutions.  In \cite{Human}, the energy coverage probability in the presence of human blockage was derived.  Discretizing harvested energy into  a finite number of power levels, the total coverage probability integrating information and energy transmission was derived by \cite{PBmmWave}.  A beam-training based mmWave WPT scheme was proposed in \cite{directional}, where the energy transmitter steers the energy beam along the direction in which it receives the strongest training signal. Considering the nonlinear behavior of energy harvesting, the coverage probability and average harvested energy were studied by \cite{nonlinear}. In \cite{Xueyuan}, the location of mmWave powered users was modeled as the Poisson cluster process and the energy and information coverage probabilities were derived. 

Most of the aforementioned works, such as \cite{mmWave_EH, howmany,Renzo_SWIPT,Human,PBmmWave,directional,nonlinear,Xueyuan}, employ the flat-top antenna pattern for mathematical tractability.  Nevertheless, it may incur great inaccuracy for evaluating  system-level performance \cite{Renzo_M, sin_antenna}.  Therefore,  some literature considers more realistic antenna patterns in the performance analysis of mmWave WPT.    In \cite{SWIPT_DA}, the microstrip patch antenna and end-fire antenna models  were used to analyze the coverage performance of SWIPT.  \cite{energybeam} adopted the Fej\'{e}r kernel model to represent the actual array antenna gain of energy transmitters. Both \cite{SWIPT_DA} and \cite{energybeam} assumed each energy transmitter is equipped with a directional antenna, while the antenna of energy receiver was assumed to be isotropic. In  \cite{mmWave_SWIPT}, both Tx and Rx were assumed to be equipped with directional antennas and the average harvested energy was derived. 

Most of the existing works on mmWave communications/WPT have assumed that the beam direction of the associated Tx-Rx pair is perfectly aligned. However, due to direction estimation error and hardware imperfection,  the beam alignment error (BAE) is always  inevitable, and affects the received signal strength greatly  \cite{IBA_sector}.  Therefore, it is necessary to investigate the performance of the mmWave transmission systems with  imperfect beam alignment (IBA). By now, much effort has been dedicated to evaluating the performance of the mmWave wireless information networks in the IBA scenario, such as   \cite{IBA_TVT,D2DIBA,IBA,sin_antenna,Real,D3}. 
 Taking into account of the BAEs in the associated Tx-Rx pair and the Interfering Txs-Rx pairs, the information coverage probability with the flat-top antenna gain model was derived by  \cite{IBA_TVT}.  With the same BAE setup, while, \cite{D2DIBA} adopted the cosine antenna and isotropic antenna patterns to model the antenna gains of the transmitters and receivers,  respectively.  In \cite{IBA},  the approximated  ergodic capacity loss of the mmWave ad hoc network with both flat-top and Gaussian antenna gain models was derived. In  \cite{sin_antenna}, the authors employed the sinc and cosine antenna patterns to derive the information coverage probability, just considering the BAEs between the typical receiver and interfering transmitters. Alternatively, using the antenna gain model suggested by the 3rd Generation Partnership Project (3GPP),  \cite{Real} investigated the impact of beam misalignment from interfering transmitters, and resorted to the curve-fitting method to derive the probability density function of the antenna gain with BAE. Furthermore, 
\cite{D3} employed both the 2D and 3D directional antenna gains and derived the information coverage probability with BAE incurred by the strongest interfering link. 

To the best of our knowledge, the impact of BAE on mmWave WPT system has not been investigated and well understood to date. Additionally, in this paper, we consider the  comprehensive IBA scenarios and the realistic 3GPP directional antenna model, which has not been well studied from the perspective of system-level performance evaluation in either mmWave communications or mmWave WPT. Specifically, there are two aspects of  our system model to be highlighted. 
First, unlike \cite{IBA_sector, IBA_TVT}, we employ the 3GPP Gaussian gain model, which can approximate the realistic directional antenna pattern exactly, to explore the impact of BAE. The reason is that the flat-top model lacks the capability of depicting the \emph{roll-off} effect of the mainlobe and thus is not suitable for accurately evaluating the impact of BAE on mmWave systems \cite{IBA,sin_antenna,Real,D2DIBA}.  Second, different from  \cite{IBA, sin_antenna,Real,D3,D2DIBA}, we consider a more realistic IBA scenario, where all Txs and Rxs are equipped with directional antennas. Moreover, not only the BAE between the associated Tx-Rx pair but also the BAEs of the non-associated Txs to the typical Rx are simultaneously taken into consideration.  Besides, the nonlinear energy harvesting model in the typical receiver is also assumed. With these models,  we derive the analytic expression of the  system-level performance of the mmWave WPT in the IBA scenario. For clarity, we summarize our main contribution as follows:
\begin{itemize}
\item  We adopt the Gaussian antenna gain model suggested by the 3GPP to represent the realistic directional antenna pattern and also assume all  transmitters and receivers are equipped with the same directional antenna.  Considering the truncated Gaussian and uniform BAE distributions, we derive the probability density functions (PDFs) of the cascaded antenna gains of a Tx-Rx pair. Taking into account of the distribution characteristics of BAE and the strong directivity of mmWave antenna, we also provide the approximated  PDFs for tractability. 
\item With the help of the Fox's H function and its series expansion, we provide a novel solution to derive the analytic expression of the energy coverage probability of the mmWave WPT in the presence of BAE. Besides,  the closed-form expression of average harvested RF energy is derived. We define the relative energy loss (REL) in order to further investigate the performance degradation incurred by BAE. 
\item Through Monte-Carlo simulations, we verify the derived energy coverage probability and average harvested RF energy. It is found that the widely used flat-top antenna gain model cannot always exactly evaluate the performance of the mmWave WPT system in the presence of BAE. We also conclude that BAE indeed lowers the energy coverage probability and the average harvested RF energy.
\end{itemize}

\emph{Notation:} $\mathbb{Z}$, $\mathbb{R}$, and $\mathbb{C}$ represent the sets of all integers, real numbers and complex numbers, respectively. $\E\{x\}$ is the expectation of random variable (r.v.) $x$ and $\Pr(A)$ is the probability of event $A$. For r.v. $x$, $f_x(\cdot)$ and $F_x(\cdot)$ stand for  the probability density function (PDF) and  cumulative distribution function (CDF) of $x$, respectively. For $x\in \mathbb{C}$, $|x|$ is the modulus  of $x$. Given $x\in \R^2$, $||x||$ means the Euclidean norm of $x$. For a complex vector $\mathbf{x}$, $\mathbf{x}^T$ and $\mathbf{x}^H$ stand for transpose and conjugate transpose of $\mathbf{x}$, respectively.  $x\sim \Gamma(k,\theta)$ means the r.v. $x$ follows Gamma distribution with PDF $f_x(x)=\frac{ x^{k-1}e^{-\frac{x}{\theta}}}{\Gamma(k)\theta^k}$ $\forall x>0, k,\theta >0$, where $\Gamma(k)$ is the Gamma function.  $\erf{x}=\frac{2}{\sqrt{\pi}}\int_0^{x} e^{-t^2} \d t$ is the Gaussian error function and $\delta(x)$ is the Dirac delta function.  $\ln x $ is the natural logarithm of $x$. For $k\leq K$ and $k, K \in \mathbb{Z}$, $\binom{K}{k}$ represents the binomial coefficient and equals to $\frac{K!}{k!(K-k)!}$. 
For r.v. $x$, $\La{x}{a}=\E\{e^{-ax}\}$ is the Laplace transform of $f_x(x)$. 

\begin{figure}[t]
	\centering
	\includegraphics[scale=0.85]{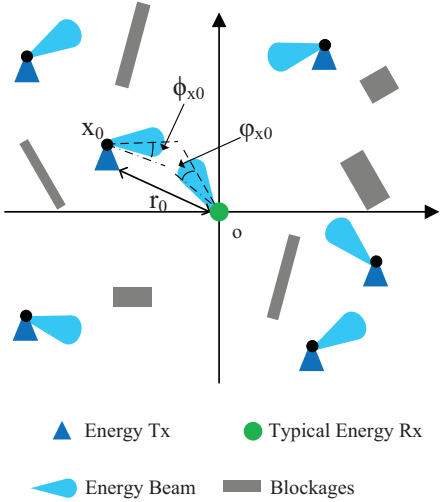}
	\caption{System model}
	\label{fig:sys}
\end{figure}
\section{System Model}
\subsection{Network and Node Models}
In this paper, we consider a mmWave wireless-powered network which consists of two types of nodes, i.e., Energy Transmitter (ETx) and Energy Receiver (ERx) (See the Fig.\ref{fig:sys}  on the next page). The ETxs are connected to stable power sources, thus having the capability of emitting  energy signals. While, the ERxs have to harvest energy to maintain their routine operation and information transmission.  The location  of all ETxs follows the homogeneous Poisson point process (HPPP) $\Phi$ with intensity $\lambda_t$ on the two-dimensional Euclidean plane. Suppose a saturated service scenario, where each ETx is assumed to serve one  dedicated ERx in a WPT block  \cite{Haenggi_book,deng_d2d}. 
We study the performance of the  typical ERx located at the origin. Let the ETx associated by the typical ERx be located at $x_0\in \R^2 $ and $||x_0||=r_0$.\footnote{Herein $r_0$ can be seen as the maximum allowable WPT distance for an  ETx-ERx pair. Accordingly, in this work we investigate the worst performance of the considered mmWave WPT system \cite{Haenggi_book,deng_d2d}.} Then, due to Slivnyak’s Theorem  \cite{Haenggi_book}, except the ETx at $x_0$, the location of other ETxs still forms a PPP with the same intensity $\lambda_t$, denoted by ${\Phi}_t=\Phi \setminus \{x_0\}$.  
Without loss of generality, the WPT duration is assumed as unit time. It means in following context, the harvested energy and the harvested power have an equivalence. 

\subsection{Channel  Model}
For mmWave links, line-of-sight (LOS) and non-line-of-sight (NLOS)  channels have sharply different propagation characteristics. Given a propagation distance $r$,  the path loss of a LOS mmWave link can be modeled as $\ell_L(r)=C_L r^{-\alpha_L}$, while in the NLOS case, the path loss is $\ell_N(r)=C_N r^{-\alpha_N}$. Here $\alpha_L$ ( $\alpha_N$) is the LOS (NLOS) path loss exponent and $C_L$ ($C_N$) represents the intercept of LOS (NLOS) link.  In general, we have  $\alpha_N>\alpha_L>0$ and $C_L \geq C_N$\cite{Renzo, blockage_model}.  Moreover, we model the small-scale fading of each mmWave link as independent Nakagami fading with parameters $m_L$ and $m_N$ for LOS and NLOS scenarios, respectively. Thus, we denote $\rho_L$ and $\rho_N$ as the small-scale channel gains for LOS and NLOS links, respectively, and then express the LOS power gain as $h_L=|\rho_L|^2\sim \Gamma(m_L,1/m_L)$ and NLOS power gain as  $h_N=|\rho_N|^2\sim \Gamma(m_N,1/m_N)$. 

The probability of a mmWave link  experiencing LOS is modeled as a function with respect to the distance from Tx to Rx \cite{5Gchannel_model}.  Several empirical and analytical blockage models had been reported in  \cite{5Gchannel_model} and \cite{blockage_model}.  Considering tractability and accuracy, we employ the random shape theory model, by which the analytical LOS probability is coincident with the empirical expression of the 3GPP blockage model \cite{blockage_model}. Hence, given the distance $r$, the LOS and the NLOS probabilities can be separately expressed as $P_L(r)=e^{-\beta r}$ and $P_{N}(r)=1-e^{-\beta r}$,  where the blockage parameter $\beta$ is determined by the average number and average perimeter of buildings in the interested area. Similar to most literature, e.g.,  [\citen{gamma_approx}]$-$[\citen{D3}], the correlation of LOS probabilities for different mmWave links is ignored.  Consequently, the ETxs in  ${\Phi}_t$ can be divided into two independent sets, namely, LOS ETx set and NLOS ETx set. Both of which follow the inhomogeneous PPP (IPPP) with intensity $P_L(r)\lambda_t$ and $P_N(r)\lambda_t$, denoted as $\Phi_L$ and $\Phi_N$, respectively.  As the ETx at $x_0$ has the distance $r_0$ to the typical ERx, the performance analysis in the LOS scenario is very similar to that in the NLOS scenario.  Furthermore, considering the harvested energy plays a crucial role in maintaining the lifetime of wireless-powered nodes, we assume the ETx only selects the ERx experiencing LOS as its target receiver like \cite{sin_antenna, deng_d2d,NY}.

\subsection{Antenna Pattern}
We assume that all ETxs and ERxs are equipped with the same directional antenna to perform mmWave beamforming. It is also assumed that all ETxs have the same transmit power $P_t$. For the typical ERx, the received RF energy transmitted by the ETx at $x$ can be given by
\begin{equation}\label{eq:GG}
\begin{split}
\varepsilon_{x,i}=P_t\ell_i(r_x)h_{x,i} {\G{}({\phi}_x)} {\G{}({\varphi}_x)}, i\in\{L,N\}
\end{split}
\end{equation}
where $\G{}(\phi_x)$ and $\G{}(\varphi_x)$ stand for the antenna gains of ETx and ERx, respectively. $\phi_x$ and $\varphi_x$ are  the orientation angles relative to the boresights of ETx and ERx, respectively, and  belong to the interval $[-\pi, \pi)$. Note that we ignore the energy harvested from the noise, as it is trivial compared with the received RF energy \cite{nonlinear}. 

In order to theoretically evaluate the system-level performance,  it is inevitable to depict the PDF of the cascaded antenna gain $\Omega_x=G({\phi}_x)G({\varphi}_x)$.  
For the uniform linear array (ULA), the Fej\'er kernel based sinc and cosine antenna patterns are employed to represent the directional antenna gain  \cite{sin_antenna,IBA_NOMA,energybeam, D2DIBA}. 
Unfortunately, it is difficult to directly derive the exact analytic expression of the PDF of  $\Omega_x$ using the sinc or cosine antenna gain model. In \cite{sin_antenna,energybeam,  D2DIBA}, the receiver was assumed to be equipped with omni-directional antenna to avoid the cascaded directional antenna gain. Although \cite{IBA_NOMA} considered a cascaded antenna gain with the sinc antenna pattern, the nodes were not stochastically deployed. 

Alternatively, the flat-top antenna gain model is mostly used in stochastic geometry coverage analysis for its mathematical tractability, e.g.,   \cite{mmWave_EH,Renzo_SWIPT,PBmmWave,howmany,IBA_sector, IBA_TVT,deng_d2d,directional,Xueyuan,Human,NY,nonlinear}. In the perfect beam alignment or slight BAE scenarios, the flat-top antenna gain model can provide the tractable theoretical expression of the system-level performance with acceptable accuracy. However,  the flat-top antenna model is lack of the capability of depicting the roll-off effect of the mainlobe and  incurs significantly inaccuracy for evaluating the system-level performance in the IBA scenario \cite{sin_antenna, IBA,IBA_J,Gaussian_1,Real}. 
It is worth mentioning that in \cite{Renzo_M} a generalized flat-top model was proposed to approximate the practical antenna gain. Nevertheless, in the IBA case, it could incur relatively high complexity to derive the PDF of the cascaded antenna gain.  

In \cite{IBA,IBA_J}, a Gaussian antenna pattern, i.e.,  $$G(\theta)=(G_m-G_s)e^{-\eta \theta^2}+G_s,$$ where $G_m$ is the maximum mainlobe gain, $G_s$ is the sidelobe gain and $\eta$ is determined by the 3dB beamwidth, was used to represent the mmWave directional antenna gain. Nevertheless, to obtain the analytic PDF of the cascaded antenna gain $\Omega_x$, the side lobe $G_s$ was ignored in the analysis of \cite{IBA,IBA_J}.  Moreover, the loss in ergodic capacity derived by \cite{IBA,IBA_J} only involves the truncated Gaussian BAE.  Differently, in the non-stochastic mmWave networks  \cite{Gaussian_1,Gaussian_2,Real} adopted the 3GPP Gaussian antenna model, which can depict the roll-off characteristics and match the measurement well. In this paper, therefore, we also employ the 3GPP Gaussian antenna model (refer to \cite{Gaussian_1}, and the references therein), i.e., 

\begin{equation}\label{eq:g_gain}
 \G{}(\theta)=\begin{cases}
G_m e^{-\eta {\theta}^2} & |\theta| \leq {\theta}_{0}, \\
 G_s &   \theta_0<|\theta|\leq \pi,
 \end{cases} 
 \end{equation} 
where $$G_m=\frac{\pi 10^{0.3\left(\frac{\theta_0^2}{\theta_{\rm 3dB}^2}\right)}}{\Theta(\theta_0,\theta_{\rm 3dB})+\pi -\theta_0}, $$
$$G_s=\frac{\pi }{ \Theta(\theta_0,\theta_{\rm 3dB})+\pi -\theta_0},$$ 
$$\Theta(\theta_0,\theta_{\rm 3dB})= \int_0^{\theta_0}10^{0.3\left(\frac{\theta_0^2-x^2}{\theta_{\rm 3dB}^2}\right)} \d x,$$ $\eta=\frac{0.3\ln 10}{\theta_{\rm 3dB}^2}$, 
$2\theta_0$ is the mainlobe (20dB) beamwidth, and $2\theta_{\rm 3dB}$ is the half-power (3dB) beamwidth.   
Since $G_s=G_me^{-\eta \theta_0^2}$, the continuity is ensured.  
According to the practical measurement reported by \cite{Gain_g,Gaussian_1}, when $\frac{\pi}{24}\leq \theta_0 \leq \frac{\pi}{6}$, $\theta_{0}$ is approximately equal to $2.6\theta_{\rm 3dB}$.  Then,  with this empirical approximation, we further obtain 
$\Theta(\theta_{0},\theta_{\rm 3dB}) = 42.6443\theta_0$, $G_m= \frac{\pi 10^{2.028}}{42.6443\theta_0+\pi}$,  and $G_s=10^{-2.028}G_m$. For convenience, we introduce the normalized antenna gain  $\widetilde{G}(\theta)=G(\theta)/G_m$, i.e., 
\begin{equation}\label{eq:ngain}
\widetilde{G}(\theta)=\begin{cases}
e^{-\eta \theta^2} & |\theta| \leq \theta_0,\\
g &  \theta_0<|\theta|\leq \pi,
\end{cases}
\end{equation} in which $g=10^{-2.028}$ and $\eta=\frac{2.028\ln10}{\theta_0^2}$. Obviously,  with the empirical expression $\theta_0=2.6  \theta_{\rm 3dB}$,  the Gaussian antenna pattern is only determined by $\theta_0$.  We herein employ \eqref{eq:ngain} to reduce the parameter number of the antenna radiation pattern. 
Besides, the cascaded normalized antenna gain is denoted by $\widetilde{\Omega}_x=\widetilde{G}(\phi_x) \widetilde{G}(\varphi_x)=\Omega_x/G^2_m$. 
\subsection{Imperfect Beam Alignment Models}
According to \eqref{eq:ngain}, if we intend to maximize the harvested energy, we can let $\phi_x = 0$ and $\varphi_x = 0$. 
In practice, however,  ${\phi}_x$ and ${\varphi}_x$ are not necessarily equal to zero due to the direction estimation error and hardware imperfection \cite{IBA_sector, IBA_TVT,D2DIBA}.  

For the associated ETx-ERx pair, as the BAE appears after the beam aligning procedure,  ${\phi}_{x_0}$ and ${\varphi}_{x_0}$ are usually modeled as independent and identically distributed truncated Gaussian random variables with zero mean and standard deviation $\sigma\geq 0$   \cite{IBA,IBA_TVT,D2DIBA}. 
Such that,  the PDF of ${\phi}_{x_0}$ can be  expressed as
\begin{equation} \label{eq:g_pdf}
f_{{\phi}_{x_0}}(\psi)=\frac{e^{-\frac{\psi^2}{2\sigma^2}}}{ \sqrt{2\pi \sigma^2}\erf{\frac{\pi}{\sqrt{2}\sigma}} }, ~~~~\psi \in [-\pi, \pi)
\end{equation}
which is named as the Gaussian BAE model. The standard deviation $\sigma$ is usually used to indicate the variability of BAE. The larger $\sigma$ is, the stronger statistical dispersion the BAE exhibits, which means the BAE becomes more severe. Observing  \eqref{eq:g_pdf}, if $\sigma \rightarrow 0$, $f_{{\phi}_{x_0}}(\psi)$ gradually converges to  $\delta(\psi)$. It is corresponding to the fact that if the beam is perfectly aligned, the  
antenna gain equals to $G_m$ with probability 1.

For the non-associated ETx-ERx pairs, an ETx in $\Phi_L$ or $\Phi_N$ just aligns its beam with the boresight of its paired ERx. To facilitate understanding and representation, for the typical ERx, given the ETx at $x \in \Phi_i$, $i\in\{L,N\}$, we can also treat ${\phi}_x$ and ${\varphi}_x$ as BAEs,  which are usually modeled as independent uniform distribution over $[-\pi, \pi)$ \cite{sin_antenna,IBA_TVT,D2DIBA}, i.e., 
\begin{equation}\label{eq:u_pdf}
f_{{\phi}_{x}}(\psi)=\frac{1}{2\pi},~~\psi \in [-\pi, \pi),~ x\in \Phi_i, ~i \in \{N,L\}.
\end{equation}
It is named as the uniform BAE model. Note that  it is assumed that BAEs at ETxs and ERxs are independently distributed  \cite{IBA,IBA_TVT, IBA_sector}. 

\subsection{Energy Harvesting Model}
 As the small-scale gains of different mmWave links are independently distributed, the harvested RF power of  the typical ERx can be written by 
 \begin{equation}
 \varepsilon_{\rm RF}=\varepsilon_{0}+\varepsilon_{L}+\varepsilon_{N}, 
 \end{equation}
 where $\varepsilon_{L}=\sum_{x\in \Phi_L}\varepsilon_{x,L}$ is the harvested power from the ETxs in $\Phi_L$, $\varepsilon_{N}=\sum_{x\in \Phi_N}\varepsilon_{x,N}$  is the harvested power from the ETxs in $\Phi_N$, and $\varepsilon_0=\varepsilon_{x_0,L}$ is the harvested power from the ETx located at $x_0$. 
 Then, the harvested direct current (DC) power at the typical ERx is 
\begin{equation}
\varepsilon_{\rm DC}=\zeta( \varepsilon_{\rm RF}).
\end{equation} 
Note that $\zeta (\varepsilon_{\rm RF})$ is the RF-DC power conversion function. In practice, $\zeta (\varepsilon_{\rm RF})$ is  a nonlinear function with respect to the input RF power $\varepsilon_{\rm RF}$ \cite{nonlinear,energybeam,practical}. Using the practical nonlinear energy harvesting model proposed in \cite{practical}, we can write the harvested DC power  as 
\begin{equation}
\varepsilon_{\rm DC}=\frac{p_m(1-\exp(-p_a \varepsilon_{\rm RF}))}{1+\exp(-p_a(\varepsilon_{\rm RF}-p_b))},
\end{equation} where $p_m$ is the maximum DC power that can be harvested by the ERx and $p_a$ and $p_b$ are the constants determined by the rectifier circuit \cite{practical}. 
 
\section{The PDFs of Antenna Gains}
In this section, we derive the PDFs of the normalized antenna gains with the BAE following the truncated Gaussian or uniform distributions. Then, the PDFs of the cascaded antenna gains with two BAE models are derived. For tractability, we also provide the  approximated PDFs of the cascaded antenna gains. 
\subsection{The PDFs of the Normalized Antenna Gains}
\begin{lemma}
If the PDF of the stochastic BAE $\psi$ is $f_{\psi}(x)$, the PDF of the normalized Gaussian antenna gain $\widetilde{G}$ is given by 
\begin{equation}
 f_{\widetilde{G}}(y)=\frac{1}{y\sqrt{-\eta\ln y }}f_{\psi}\left(\sqrt{\frac{-\ln y}{\eta}}\right)+(1-P_0) \delta(y-g),
 \end{equation} where $y\in [g,1]$ and  $P_0=\Pr(|\psi| \leq \theta_0)=\int_{-\theta_0}^{\theta_0}f_{\psi}(x)\d x$.
\end{lemma}

\begin{IEEEproof}
See Appendix A. 
\end{IEEEproof}

Based on the Lemma 1,  we have following two corollaries according to \eqref{eq:g_pdf} and \eqref{eq:u_pdf}. 
\begin{corollary}
If the BAE $\psi$ follows the truncated Gaussian distribution with zero mean and variance $\sigma^2$ over $[-\pi, \pi)$, the PDF of $\widetilde{G}$  can be expressed as 
\begin{equation}\label{eq:g_g}
f_{\widetilde{G}}(y)=\frac{y^{\frac{1}{2\eta \sigma^2}-1}}{\sqrt{2\pi \eta \sigma^2} \erf{\frac{\pi}{\sqrt{2\sigma^2}}}\sqrt{-\ln y}}+(1-P_{0}^G)\delta(y-g),
\end{equation} where $y\in [g,1]$ and $P_{0}^G=\int_{-\theta_{0}}^{\theta_{0}}f_{\phi_{x_0}}(\psi)\d \psi =\erf{\frac{\theta_0}{\sqrt{2\sigma^2}}}/\erf{\frac{\pi}{\sqrt{2\sigma^2}}}$. 
\end{corollary}
 Thus, for the associated ETx-ERx pair, the PDF of the normalized antenna gain of ERx or ETx is equal to  \eqref{eq:g_g}. 
\begin{corollary}
If the BAE $\psi$ follows the uniform distribution over $[-\pi, \pi)$, the PDF of $\widetilde{G}$  can be written by
\begin{equation}\label{eq:g_u}
f_{\widetilde{G}}(y)=\frac{1}{2\pi y\sqrt{-\eta \ln y}}+(1-P_{0}^U)\delta(y-g),
\end{equation} where $y\in [g,1]$ and  $P_{0}^U=\int_{-\theta_{0}}^{\theta_0}f_{\phi_x}(\psi)\d \psi=\frac{\theta_0}{\pi}$.
\end{corollary}
For the non-associated ETx-ERx pairs, the PDF of involved antenna gains is illustrated by the Corollary 2. 

\subsection{The PDFs of the Cascaded Antenna Gains}
By \eqref{eq:GG}, the cascaded antenna gain $\Omega_x$ plays a crucial role in the system performance. Herein, we intend to derive the PDFs of $\widetilde{\Omega}_x$ with both BAE models respectively. To this end, we have following two theorems. 
\begin{theorem}
The PDF of  the cascaded normalized antenna gain $\widetilde{\Omega}_x=\widetilde{G}(\phi_x)\widetilde{G}(\varphi_x)$ with truncated Gaussian BAE model can be written as \eqref{eq:pdf_g} at the top of the next page. 
\begin{figure*}[!t]
\normalsize
\setcounter{equation}{11}
\begin{equation}\label{eq:pdf_g}
f_{\widetilde{\Omega}_x}(\Omega)=
\begin{cases}
\frac{\Omega^{\frac{1}{2\eta \sigma^2}-1}}{2\eta \sigma^2 {\rm erf}^2\left(\frac{\pi}{\sqrt{2\sigma^2}}\right)}& \Omega \in [g,1]\\
\frac{\Omega^{\frac{1}{2\eta \sigma^2}-1}}{\pi \eta \sigma^2 {\rm erf}^2\left(\frac{\pi}{\sqrt{2\sigma^2}}\right)}\arctan \left(\frac{\ln \Omega-2\ln g}{2\sqrt{\ln g \ln \frac{\Omega}{g}}}\right)+\frac{2(1-P_{0}^G)g^{-\frac{1}{2\eta \sigma^2}}}{\sqrt{2\pi \eta\sigma^2}\erf{\frac{\pi}{\sqrt{2\sigma^2}}}}\frac{\Omega^{\frac{1}{2\eta \sigma^2}-1}}{\sqrt{-\ln \frac{\Omega}{g}}}+\delta(\Omega-g^2)(1-P_{0}^G)^2& \Omega \in [g^2,g)
\end{cases}
\end{equation}
\vspace*{4pt}
\end{figure*}
\end{theorem}
\begin{IEEEproof}
See Appendix B
\end{IEEEproof}

\begin{theorem}
The PDF of the cascaded normalized antenna gain $\widetilde{\Omega}_x=\widetilde{G}(\phi_x)\widetilde{G}(\varphi_x)$ with uniform BAE model can be written as \eqref{eq:pdf_u} at the top of the next page. 
\end{theorem}
\begin{figure*}
\normalsize
\setcounter{equation}{12}
\begin{equation}\label{eq:pdf_u}
f_{\widetilde{\Omega}_x}(\Omega)=
\begin{cases}
\frac{1}{4\pi \eta \Omega} & \Omega\in [g,1]\\
\frac{1}{2\pi^2 \eta \Omega}\arctan\left(\frac{\ln\Omega-2\ln g}{2\sqrt{\ln g \ln \frac{\Omega}{g}}}\right) +\frac{1-P_{0}^U}{\Omega\pi \sqrt{\eta}}\frac{1}{\sqrt{-\ln \frac{\Omega}{g}}}+(1-P_{0}^U)^2\delta(\Omega-g^2) & \Omega \in [g^2,g)
\end{cases}
\end{equation}
\setcounter{equation}{13}
\hrulefill
\vspace*{4pt}
\end{figure*}
\begin{IEEEproof}
The proof  of Theorem 2 is similar to that of Theorem 1, therefore, we omit the proof for clarity. 
\end{IEEEproof}
\subsection{The Approximated PDFs of the Cascaded Antenna Gains}

Although Theorem 1 and 2 show the PDFs of $\widetilde\Omega$ with the Gaussian and uniform BAE models respectively, the arctan functions in \eqref{eq:pdf_g} and \eqref{eq:pdf_u} make further  analysis less tractable. Therefore, we provide two approximated PDFs, by considering the distribution characteristics of BAE models and the strong directivity of mmWave antenna.    
\subsubsection{The Approximated PDF With the Gaussian BAE model}
In the mmWave WPT system, the associated ETx-ERx pair is expected to employ  elaborately designed beam alignment algorithms, such as \cite{CBA}, to minimize  $|\phi_{x_0}|$ and $|\varphi_{x_0}|$ as much as possible. It is therefore reasonable to infer that  $|\phi_{x_0}|$ and $|\varphi_{x_0}|$ are far less than $\theta_0$ in most cases.  For instance,  it is straightforwardly assumed by \cite{robust} that   $\phi_{x_0}$ and $\varphi_{x_0}$ 
 lie in $[-\theta_0, \theta_0]$. 
The authors of \cite{IBA} and \cite{IBA_J} ignored the cascaded antenna gain involving the sidelobe gain, considering it has the relatively small value and happens in a very low probability in the Gaussian BAE scenario. Following these works,  we also ignore the sidelobe gain in the cascaded antenna gain, i.e., the component of \eqref{eq:pdf_g} over $[g^2,g)$. As a result, $f_{\widetilde \Omega_{x_0}}(\Omega)$ with the Gaussian BAE model can be approximately  presented as
\begin{equation}
\label{eq:apdf_g}
f_{\widetilde\Omega_{x_0}}(\Omega)\sim \frac{\Omega^{\frac{1}{2\eta \sigma^2}-1}}{2\eta\sigma^2 {\rm erf}^2{\left(\frac{\pi}{\sqrt{2\sigma^2}}\right)}}, ~~~\Omega\in [g,1].
\end{equation}
When $\sigma= 0$, we have $f_{\widetilde \Omega_{x_0}}(\Omega) = \delta(\Omega-1)$. 
\subsubsection{The Approximated PDF With the Uniform BAE model}
Recalling the expression of $G_m$, the strong directional antenna means $\theta_0$ is far less than $\pi$. Hence,  in the uniform BAE case,  the event that  $\widetilde \Omega_x$ equals to the product of two mainlobes occurs in an extremely small probability, i.e., $\theta_0^2/\pi^2\ll 1$. Furthermore, by  \eqref{eq:f1} in the Appendix B, it  can be inferred  that the arctan term in \eqref{eq:pdf_u} is generated by the product of two mainlobes. Therefore, it is reasonable to neglect the arctan term of \eqref{eq:pdf_u} in the uniform BAE case. Then, we can approximate \eqref{eq:pdf_u} by $f_{\widetilde\Omega_x}(\Omega)\sim $
\begin{equation}\small
\label{eq:apdf_u}
\begin{cases}
\frac{1-P_0^U}{\pi \Omega\sqrt{\eta}\sqrt{-\ln \frac{\Omega}{g}}}+\delta(\Omega-g^2)(1-P_0^U)^2,&\Omega\in [g^2,g) \\
\frac{1}{4\pi \eta \Omega}, & \Omega \in [g,1] 
\end{cases}
\end{equation} 

\begin{figure}[!tb]
\centering
\includegraphics[scale=0.42]{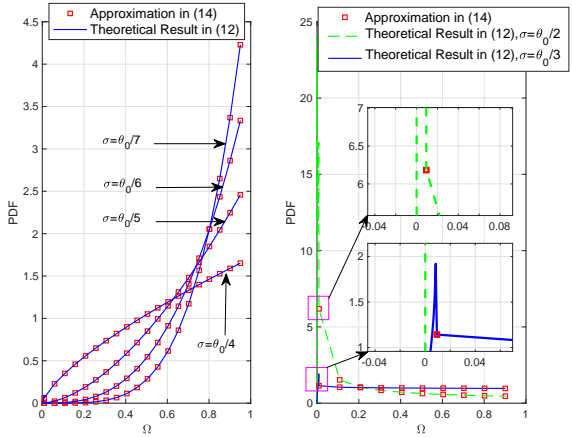}
\caption{The PDF of $\widetilde\Omega_{x_0}$ with truncated  Gaussian error, $\theta_0=\frac{\pi}{12}$ .}
\label{fig:G_f}
\end{figure}
\begin{figure}[!tb]
\centering
\includegraphics[scale=0.42]{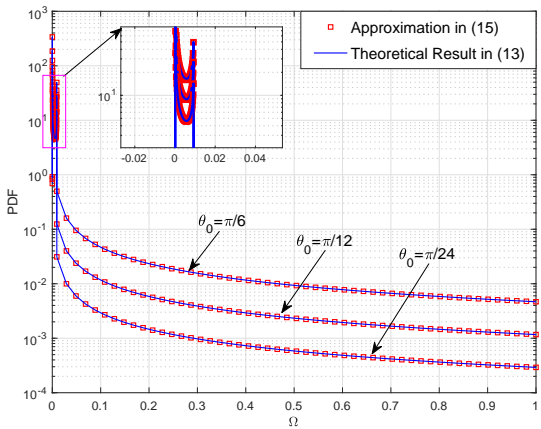}
\caption{The PDF of $\widetilde \Omega_x$ with uniform error.}
\label{fig:U_f}
\end{figure}
\subsubsection{Verification of the Approximated PDFs}
To verify our approximations, we draw $f_{\widetilde \Omega}(\Omega)$ with Gaussian and  uniform BAE models in \figref{fig:G_f} and \figref{fig:U_f}, respectively.  For the Gaussian BAE model, we draw the PDF of $\widetilde G(\theta)$ with $\theta_0=\frac{\pi}{12}$ as an example to verify our  approximation.\footnote{$\forall \theta_0 \in [\frac{\pi}{24}, \frac{\pi}{6}]$, the approximated PDF of  $\widetilde\Omega_{x_0}$ shows the similar accuracy with various $\sigma$.} From the left subfigure of \figref{fig:G_f},  we can see that with $\sigma=\theta_0/7,\theta_0/6, \theta_0/5,$ and $\theta_0/4$, the approximated PDFs match the theoretical PDFs closely. As $\sigma$ decreases, the PDF of $\widetilde{\Omega}_{x_0}$ tends to be a pulse-like  function.  We can infer that it asymptotically converges to $\delta(\Omega-1)$ as $\sigma\rightarrow 0$.  While, in the right subfigure of  \figref{fig:G_f}, for $\sigma=\theta_0/3$, there is a slight difference between the results of (12) and (14). It is because the probability that the sidelobe appears in the cascaded antenna gain becomes larger when $\sigma$ grows.  Thus, with $\sigma=\theta_0/2$, the difference generated by our approximation seems extremely apparent. Accordingly, if $\sigma \le \theta_0/3$,   (14) is an appropriate approximation of (12). \footnote{ Given $\sigma =\frac{\theta_0}{3}$,  there is always $P_0^G(\theta_0) \approx 0.9973$ for $\theta_0\in [\frac{\pi}{24}, \frac{\pi}{6}]$. While, given $\sigma=\frac{\theta_0}{2}$, we have  $P_0^G(\theta_0) \approx 0.9545$ for $\theta_0\in [\frac{\pi}{24}, \frac{\pi}{6}]$. Consequently, the probabilities of two mainlobes cascading with $\sigma=\frac{\theta_0}{3}$ and $\sigma=\frac{\theta_0}{2}$, i.e., $P_0^G(\theta_0)\cdot P_0^G(\theta_0)$, are about 0.9946 and 0.9111, respectively. Apparently, for $\sigma=\frac{\theta_0}{3}$,  $\forall\theta_0\in [\frac{\pi}{24}, \frac{\pi}{6}]$, there is $P_0^G(\theta_0)\cdot P_0^G(\theta_0)\approx 1$. That is why we here choose $\sigma \leq \theta_{0}/3$ as the approximation condition. Note that it is weaker than the assumption that $\phi_{x_0}, \psi_{x_0} \in[-\theta_0,\theta_0]$ adopted by \cite{robust}.} 
Moreover, from  \figref{fig:U_f} we can see that the curves of (15) approach those of (13) extremely closely. 
Summarily, it is verified that both approximated PDFs can be used to analyze the system-level performance under our considered circumstance. 

\section{Energy Coverage Analysis}
In this section, we focus on analyzing the energy coverage probability of the typical ERx. Energy coverage probability is defined as the probability that the harvested DC energy is larger than a pre-defined threshold, which is always the minimum required energy for information transmission or other operations. 

In \cite{Gfunc}, the Meijer G-function was used to derive the analytic expression of information coverage probability of mmWave transmission. Theoretically speaking, we can also adopt this successful approach in our analysis. By \cite{ISSF}, however, to obtain the analytic expression in the form of the Meijer G-function, the path loss exponents should be positive integers, which limits the application of the Meijer G-function based method. Alternatively, in this paper, without path loss exponent limitation,  we provide an  analytic expression of energy coverage probability with the help of  Fox's H function.\footnote{The Fox's H function is a general function which can encompass almost all commonly used functions, e.g., Meijer G-function. Although the Fox's H function is defined by an integral in a nonanalytic form, like the widely used Gamma function, Q-function, Hypergeometric function, Meijer G-function, etc.,  a look-up-table (LUP) storing the values of Fox's H function can be generated via numerical methods. A Matlab program for evaluating the  Fox's H function was provided in \cite{H_Thesis}.  More details about the Fox's H function can be found in  \cite{hfunc}.}

\begin{table}
	\caption{Definitions of Variables in Section IV}
	\label{table:def}
	\centering
	\begin{spacing}{1.5}
	\begin{tabular}{|l|}
		\hline
		
		$\tilde{\varepsilon}_{\rm th}= -\frac{1}{p_a}\ln \left( \frac{p_m-\varepsilon_{th}}{p_m+\varepsilon_{th}\exp(p_ap_b)}\right)$ \\ 
		\hline 
		
		$A=K(K!)^{-1/K}$\\ 
		\hline 
	
		$a_k=\frac{Ak}{\tilde{\varepsilon}_{th}}$\\ 
		\hline 
	
	$\mathfrak{F}(x)= x^{\varpi}{ _2}\F_1 (m_L, \varpi;1+\varpi;-\gamma_k x)$ \\ 
		\hline
	
		$\varpi=\frac{1}{2\eta \sigma^2}$\\
		\hline
		$\gamma_k=\frac{a_k P_tG_m^2C_Lr_0^{-\alpha_L}}{m_L}$\\
		\hline
		$\omega_{x,L}=\widetilde \Omega_x h_{x,L}$, $x\in \Phi_L$\\
		 \hline
		  ${\chi}_{L,z} =\E\{\omega_L^{z}\}=\E \{\widetilde \Omega^z\}\E\{h_L^z\}$, $z\in \R^{+}$\\
		  \hline
		  $\chi_{N,z}=\E\{\omega_N^{z}\}=\E \{\widetilde \Omega^z\}\E\{h_N^z\}$, $z\in \R^{+}$\\
		  \hline 
	\end{tabular} 
\end{spacing}
\end{table}
Letting $\varepsilon_{\rm th}$ be the DC energy threshold, we can write the energy coverage probability of  the typical ERx as   
\begin{equation}
P_{ec}=\Pr(\varepsilon_{\rm DC}>\varepsilon_{\rm th} )=\Pr(\varepsilon_{\rm RF} > \tilde{\varepsilon}_{\rm th}), 
\end{equation} 
where $\tilde{\varepsilon}_{\rm th}= -\frac{1}{p_a}\ln \left( \frac{p_m-\varepsilon_{th}}{p_m+\varepsilon_{th}\exp(p_ap_b)}\right)$ is the equivalent RF energy threshold. By the expression of  $\tilde{\varepsilon}_{\rm th}$, If $\varepsilon_{\rm th}\geq p_m$, there is $P_{ec}=0$. Then, we let $\varepsilon_{th} < p_m$ to investigate the performance of $P_{ec}$. Note that $\varepsilon_0$, $\varepsilon_L$ and $\varepsilon_N$ are independent of each other.  
Following the widely adopted Gamma r.v.  approximation \cite{mmWave_EH,nonlinear}, i.e., using $\mu_K \sim \Gamma (K, \frac{1}{K})$, $K\in \mathbb{Z}^{+}$, instead of  $1$,  we can rewrite $P_{ec}$ as
\begin{equation}
\begin{split}
P_{ec}&\approx\E_{\varepsilon_0,\varepsilon_L,\varepsilon_N}\left\{\Pr\left(\mu_{K} < \frac{\varepsilon_0+\varepsilon_L+\varepsilon_N}{\tilde{\varepsilon}_{th}}\right)\right\}\\
&\overset{(a)}{\approx} \E_{\varepsilon_0,\varepsilon_L,\varepsilon_N}\left\{\left[1-\exp \left(-\frac{A(\varepsilon_0+\varepsilon_L+\varepsilon_N)}{\tilde \varepsilon_{th}}\right)\right]^K\right\},\\
&=\sum_{k=0}^K (-1)^k \binom{K}{k}\La{\varepsilon_0}{a_k} 
\La{\varepsilon_L}{a_k}\La{\varepsilon_N}{a_k}, 
k\in\mathbb{Z}^{+} \label{eq:ecp}
\end{split}
\end{equation}
in which  $a_k=\frac{Ak}{\tilde{\varepsilon}_{th}}$  and $A=K(K!)^{-1/K}$. (a) is based on Lemma 5 in \cite{mmWave_EH}. 
Next, we  derive the analytic expressions of $\La{\varepsilon_0}{a_k}$, $  
\La{\varepsilon_L}{a_k}$, and $\La{\varepsilon_N}{a_k}$, respectively. 
\subsection{The Analytic Expression of $\La{\varepsilon_0}{a_k}$}
As for $\varepsilon_0$,  $h_{x_0,L}$ and $\widetilde\Omega_{x_0}$ are independent random variables, we have
\begin{equation}\small
\begin{split}
&\La{\varepsilon_0}{a_k}=\E_{\widetilde \Omega_{x_0},h_{x_0,L}}\left\{ e^{-a_kP_tG_m^2 \ell_L(r_0)\widetilde\Omega_{x_0} h_{x_0,L}}\right\}\\
&=\E_{\widetilde \Omega_{x_0}}\left\{\left(1+\frac{a_k P_tG_m^2 C_Lr_0^{-\alpha_L}\widetilde \Omega_{x_0}}{m_L}\right)^{-m_L}\right\}\\
&\approx \int_{g}^1 \left(1+\frac{a_k P_tG_m^2 C_Lr_0^{-\alpha_L}\widetilde \Omega_{x_0}}{m_L}\right)^{-m_L} \frac{\widetilde \Omega_{x_0}^{\frac{1}{2\eta\sigma^2}-1}}{2\eta \sigma^2 {\rm erf}^2\left({\frac{\pi}{\sqrt{2\sigma^2}}}\right)} \d\widetilde \Omega_{x_0} \\
&\overset{(b)}{=}\frac{\mathfrak{F}(1)-\mathfrak{F}(g)}{ {\rm erf}^2\left({\frac{\pi}{\sqrt{2\sigma^2}}}\right)} 
\end{split}
\end{equation} 
where $\mathfrak{F}(x) \triangleq x^{\varpi}{ _2}\F_1 (m_L, \varpi;1+\varpi;-\gamma_k x)$, $\varpi=\frac{1}{2\eta \sigma^2}$, $\gamma_k=\frac{a_k P_tG_m^2C_Lr_0^{-\alpha_L}}{m_L}$, and $_2\F_1(a,b;c;z)$ is the Gauss hypergeometric function \cite{ISSF}. Note that in $(b)$ we resort to the following equation \cite{Math_2F1}, 
\begin{equation*}
\begin{split}
\int_0^1 b (1+z t)^{-a} t^{b-1} \d t=&{_2}\F_1 (a,b;1+b;-z),\\
& \forall z\geq -1, a>0, b>0, z, a, b\in\mathbb{R}.
\end{split}
\end{equation*}
When $\sigma=0$, i.e., the perfect beam alignment scenario, we can easily obtain 
\begin{equation}
\La{\varepsilon_0}{a_k}=\left(1+\frac{a_k P_tG_m^2 C_Lr_0^{-\alpha_L}}{m_L}\right)^{-m_L}.
\end{equation}
\subsection{The Analytic Expression of $\La{\varepsilon_L}{a_k}$}
Define $\omega_{x,L}\triangleq\widetilde \Omega_x h_{x,L}$, $x\in \Phi_L$. 
As $\Phi_L$ follows the IPPP with intensity $P_L(r_x)\lambda_t$, the 2-tuple  $\{\omega_{x,L}\}\times\Phi_L$ forms a marked IPPP (MIPPP). Due to the probability generating functional (PGFL) of MIPPP \cite{Haenggi_book}, we have
\begin{equation}\label{eq:laplace_L}
\begin{split}
&\La{\varepsilon_L}{ a_k }=\E_{\Phi_L,\omega_{x,L}}\left\{ \prod_{x\in \Phi_L}e^{-a_k P_tG_m^2 C_Lr^{-\alpha_L}_x \omega_{x,L}}\right\}\\
&=e^{-2\pi \lambda_t \int_0^{\infty}\left(1-\E_{\omega_L}\left\{e^{-a_k P_t G_m^2 C_L r^{-\alpha_L}\omega_L}\right\}\right)e^{-\beta r}r \d r }\\
&\overset{(c)}{=}e^{-2\pi \lambda_t\left( \frac{1}{\beta^2}-\E_{\omega_L}\left\{\int_0^{\infty} e^{-\beta r - a_k P_t G_m^2C_L r^{-\alpha_L}{\omega_L}}r \d r\right\} \right)}.
\end{split} 
\end{equation} In $(c)$, due to the Fubini's theorem \cite{Fub}, we exchange the order of integral and expectation operations. 
Due to  [\citen{mathai_hfunc}, (1.9.5)],  we have  
\begin{equation}\label{eq:hfunc}
\begin{split}
&\int_0^{\infty} e^{-\beta r - a_k P_tG_m^2C_L r^{-\alpha_L}{\omega_L}}r \d r \\
&=\frac{1}{\alpha_L\beta^2}{\H}_{0,2}^{2,0}\left[\beta\left(a_kP_t G_m^2 C_L\omega_L\right)^{\frac{1}{\alpha_L}}\Big|  _{(2,1)(0,\frac{1}{\alpha_L})}\right].
\end{split}
\end{equation}
Note that $\H_{0,2}^{2,0}[\cdot]$ is the Fox's H function and is defined by \cite{hfunc,mathai_hfunc}. Consequently, there is $\La{\varepsilon_L}{ a_k }=$
\begin{equation}
e^{-2\pi \lambda_t\left( \frac{1}{\beta^2}-\frac{1}{\alpha_L\beta^2}\E_{\omega_L}\left\{{\H}_{0,2}^{2,0}\left[\beta(a_k P_t G_m^2 C_L\omega_L )^{\frac{1}{\alpha_L}}\bigg|  _{(2,1)(0,\frac{1}{\alpha_L})}\right]\right\} \right)}.
\end{equation}

Before solving the expectation  of  $\H_{0,2}^{2,0}[\cdot]$ with respect to $\omega_L$, we introduce the Lemma 2 as follows. 

\begin{lemma}
For $t\in \mathbb{Z}^{+}$, there is
\begin{equation*}
\begin{split}
&{\H}_{p,q}^{m,n}\left[xy\Big|_{(b_1,B_1)\cdots (b_q,B_q)}^{(a_1,A_1)\cdots(a_p,A_p)} \right]=\\
& x^{\frac{b_1}{B_1}}\sum_{t=0}^{\infty}\frac{\left(1-x^{\frac{1}{B_1}}\right)^{t}}{t!}{\H}_{p,q}^{m,n}\left[y\Big| _{(t+b_1,B_1)(b_2,B_2)\cdots (b_q,B_q)}^{(a_1,A_1)\cdots(a_p,A_p)}\right].
\end{split}
\end{equation*}
\end{lemma}
\begin{IEEEproof}
See [\citen{hfunc}, (1.88)].
\end{IEEEproof}
By Lemma 2, we can further attain
\begin{equation*}\small
\begin{split}
&{\H}_{0,2}^{2,0}\left[\beta\left(a_k P_tG_m^2 C_L\omega_L\right)^{\frac{1}{\alpha_L}}\bigg|  _{(2,1)(0,\frac{1}{\alpha_L})}\right]\\
&=\omega_L^{\frac{2}{\alpha_L}} \sum_{t=0}^{\infty} \frac{\left(1-\omega_L^{\frac{1}{\alpha_L}}\right)^t}{t!}{\H}_{0,2}^{2,0} \left[\beta (a_k P_t G_m^2 C_L)^{\frac{1}{\alpha_L}}\bigg|_{(t+2,1)(0,\frac{1}{\alpha_L})}\right]\\
&
\overset{(d)}{=}\sum_{t=0}^{\infty}\sum_{q=0}^{t}\frac{(-1)^q \binom{t}{q}\omega_L^{\frac{q+2}{\alpha_L}}}{t!}{\H}_{0,2}^{2,0}\left[\beta(a_k P_t G_m^2 C_L)^{\frac{1}{\alpha_L}}\bigg|_{(t+2,1)(0,\frac{1}{\alpha_L})}\right]. 
\end{split}
\end{equation*} 
In $(d)$, we apply Binomial theorem. Hence, there is
\begin{equation}\small \label{eq:LeL}
\begin{split}
&\La{\varepsilon_L}{ a_k }\approx \exp\left\{-2\pi \lambda_t\left[ \frac{1}{\beta^2}-\frac{1}{\alpha_L\beta^2} \sum_{t=0}^{\infty}\sum_{q=0}^{t}\frac{(-1)^q \binom{t}{q}{\chi}_{L,{\frac{q+2}{\alpha_L}}}}{t!}\right.\right.\\
&\left.\left. {\H}_{0,2}^{2,0}\left[\beta(a_k P_tG_m^2C_L)^{\frac{1}{\alpha_L}}\bigg|_{(t+2,1)(0,\frac{1}{\alpha_L})}\right] \right]\right\},
\end{split}
\end{equation}
where ${\chi}_{L,z} \triangleq\E\{\omega_L^{z}\}=\E\{\widetilde \Omega^z\}\E\{h_L^{z}\}$, $z \in \R^{+}$. 

\begin{proposition}
If the PDF of  $\widetilde{\Omega}$  follows  \eqref{eq:apdf_u}, $\forall z \in \mathbb{R}^{+}$, there is 
\begin{equation}
\begin{split}
\E\{\widetilde \Omega^{z}\}= &\frac{1-P_0^U}{\pi \sqrt{\eta}}\frac{g^z \sqrt{\pi}\erf{\sqrt{-z\ln g}}}{\sqrt{z}}+\frac{1-g^z}{4\pi \eta z}  \\
&+(1-P_0^U)^2g^{2z}  , 
\end{split}
\end{equation}
\end{proposition}

\begin{IEEEproof}
By \eqref{eq:apdf_u}, it is straightforward to obtain Proposition 1. The detailed proof is omitted to save space.
\end{IEEEproof}
\begin{proposition}
If  $h\sim \Gamma(m,1/m)$, $m\in\mathbb{Z}^{+}$,  $\forall z \in \mathbb{R}^{+}$, there is
\begin{equation}
\E\{h^z\}=\frac{\Gamma(m+z)}{\Gamma(m)m^{z}}, 
\end{equation}
\end{proposition}
\begin{IEEEproof}
By the PDF of $h$, it is  conveniently to prove  Proposition 2. 
\end{IEEEproof}
Therefore, substituting  Proposition 1 and 2 into \eqref{eq:LeL}, we can obtain the analytic expression of $\La{\varepsilon_L}{ a_k }$. 

\subsection{The Analytic Expression of $\La{\varepsilon_N}{a_k}$}

Similarly to \eqref{eq:laplace_L},  defining $\omega_{x,N}=\widetilde\Omega_x h_{x,N}$, we have 
\begin{equation}\small
\begin{split}
&\La{\varepsilon_N}{a_k}=\E_{\Phi_N,\omega_{x,N}}\left\{\prod_{x\in \Phi_N}e^{-a_k P_t G_m^2 C_Nr^{-\alpha_N}_{x} \omega_{x,N}} \right\}\\
\\
&= e^{-2\pi \lambda_t \left(\int_0^{\infty}\left(1-\E_{\omega_N}\left\{e^{-a_k P_t G_m^2 C_N r^{-\alpha_N} \omega_N}\right\}\right)(1-e^{-\beta r})r \d r\right) }\\
&=\exp \left\{-2\pi \lambda_t \left[\underbrace{\int_0^{\infty}\left(1- \E_{\omega_N}\left\{e^{-a_k P_t G_m^2  C_N r^{-\alpha_N}\omega_N}\right\}\right)r \d r}_{\mathscr{L}_1(a_k)} \right.\right.\\
&~~\left.\left.-\underbrace{\int_0^{\infty} \left(1-\E_{\omega_N}\left\{e^{-a_k P_t G_m^2 C_N r^{-\alpha_N}\omega_N}\right\}\right)e^{-\beta r}r \d r}_{\mathscr{L}_2(a_k)}\right]\right\}.
\end{split}
\end{equation}
Next, we derive the expressions of $\mathscr L_1(\omega_N)$ and $\mathscr L_2(\omega_N)$ separately.

Firstly, after some manipulations, it is easy to know
\begin{equation}
\mathscr L_1(a_k)=\frac{1}{2}(a_k P_t G_m^2C_N)^{\frac{2}{\alpha_N}}\chi_{N,\frac{2}{\alpha_N}} \Gamma\left(1-\frac{2}{\alpha_N}\right), 
\end{equation} where $\chi_{N,z}\triangleq \E\{\omega_N^{z}\}= \E \{\widetilde \Omega^z\}\E\{h_N^z\}$, $z\in \R^{+}$, which can be also written in the analytic form due to  Proposition 1 and 2. Following the derivations from \eqref{eq:hfunc} to \eqref{eq:LeL}, similarly, we can obtain
\begin{equation}\small
\begin{split}
\mathscr{L}_2(a_k)\approx &\frac{1}{\beta^2}-\frac{1}{\alpha_N\beta^2}
\sum_{t=0}^{\infty}\sum_{q=0}^{t}\frac{(-1)^q \binom{t}{q}{\chi}_{N,\frac{q+2}{\alpha_N}}}{t!}\times\\&{\H}_{0,2}^{2,0}\left[\beta(a_k P_t G_m^2 C_N)^{\frac{1}{\alpha_N}}\bigg|_{(t+2,1)(0,\frac{1}{\alpha_N})}\right]. 
\end{split}
\end{equation} 
Consequently,  the analytic expression of $\La{\varepsilon_N}{a_k}$ is also derived.
Therefore, we can get the analytic expression of $P_{ec}$ with $\La{\varepsilon_0}{a_k}$, $\La{\varepsilon_L}{a_k}$,  and $\La{\varepsilon_N}{a_k}$.  

 It is worth noting that for applying Proposition 1 and 2, we need to let $1/\alpha_L \in \mathbb{R}^{+}, 1/\alpha_N \in \mathbb{R}^{+}$. Clearly, it always holds for the practical condition $\alpha_L> \alpha_N >0$. So, we put no limitation on the path loss exponents. In addition, for clarity, we summarized all newly-defined variables in Section IV in Table \ref{table:def}.   

\section{Average Harvested Energy}
Although the derived $P_{ec}$ can be used to evaluate the energy coverage performance of mmWave WPT, it may not provide explicit and direct insight into the effect of BAE. Thus, in this section, the average harvested energy is addressed to further investigate the effect of BAE. 

By \cite{mmWave_EH,nonlinear}, the average harvested DC energy can be expressed as
\begin{equation}
{\varepsilon}_{\rm avg}=\varepsilon_{\rm min} P_{\rm ec}(\varepsilon_{\rm min} )+\int_{\varepsilon_{\rm min} }^{\infty} P_{\rm ec}(\varepsilon) \d \varepsilon, 
\end{equation} where $\varepsilon_{\rm min}$ is the minimum energy threshold. Apparently, it is extremely difficult to give an analytic expression of ${\varepsilon}_{\rm avg}$ because of the complicated expression of energy coverage probability. 

To achieve the closed-form  result, we herein consider the linear energy harvesting (EH) model, i.e., $ \zeta(\varepsilon_{\rm RF})=\zeta \varepsilon_{\rm RF}$ like  \cite{mmWave_SWIPT, PBmmWave,howmany,directional,Xueyuan}.  Specifically, we set $\zeta=1$ as \cite{HowmanyRF}, which means we investigate the  performance of average harvested RF energy at the typical ERx.  
With this linear EH model, we have
\begin{equation}\label{eq:ae}
\varepsilon_{\rm avg}=\E\{\varepsilon_0\}+\E\{\varepsilon_L\}+\E\{\varepsilon_N\}.
\end{equation}
Obviously, there is
\begin{equation}\label{eq:e0}
\begin{split}
\E\{\varepsilon_0\}&=P_tG_m^2C_L r_0^{-\alpha_L}\E\{\widetilde\Omega_{x_0}\}\E\{h_{x_0,L}\}\\
&\approx P_tG_m^2C_Lr_0^{-\alpha_L}\frac{1-g^{\frac{1}{2\eta\sigma^2}+1}}{(2\eta\sigma^2+1){\rm erf}^2\left(\frac{\pi}{\sqrt{2\sigma^2}}\right)}
\end{split}.
\end{equation}
As for $\varepsilon_L$, 
due to Campell's Theorem \cite{Haenggi_book}, we have
\begin{equation}\label{eq:eL}
\begin{split}
&\E\{\varepsilon_L\}=\E_{\Phi_L,\omega_{x,L}}\left\{\sum_{x\in \Phi_L} P_tG_m^2\omega_{x,L} C_L r_{x,L}^{-\alpha_L}\right\}\\
&=2\pi \lambda_t\int_0^{\infty}\E_{\omega_{L}}\left\{P_tG_m^2\omega_{L}C_L r_{L}^{-\alpha_L}\right\}e^{-\beta r}r \d r\\
&\overset{(e)}{\approx} 2\pi \lambda_t P_tG_m^2C_L\chi_{L,1}\int_1^{\infty} r^{-\alpha_L} e^{-\beta r}r\d r\\
&=2\pi \lambda_t P_tG_m^2C_L\chi_{L,1} \beta^{\frac{\alpha_L-1}{2}-1}e^{-\frac{\beta}{2}}\W_{-\frac{\alpha_L-1}{2},\frac{2-\alpha_L}{2}}(\beta)
\end{split},
\end{equation} where $\W_{a,b}(x)$ is the Whittaker W function [\citen{table}, (3.381.6)] and can be efficiently calculated by Matlab. 
To avoid the singularity incurred by the simplified path loss model \cite{Haenggi_book}, in $(e)$ we only consider the far field energy signals.  
In the same way, we obtain
\begin{equation}\label{eq:eN}
\begin{split}
&\E\{\varepsilon_N\}=\E_{\Phi_N,\omega_{x,N}}\left\{\sum_{x\in \Phi_N} P_t G_m^2 \omega_{x,N} C_N r_{x,N}^{-\alpha_N}\right\}\\
&\approx 2\pi \lambda_t\int_1^{\infty}\E_{\omega_{N}}\left\{P_t G_m^2 \omega_{N} C_N r_{N}^{-\alpha_N}\right\} (1-e^{-\beta r})r \d r\\
&=2\pi \lambda_t P_t G_m^2  C_N\chi_{N,1}\times\\&\left(\frac{1}{\alpha_N-2}-\beta^{\frac{\alpha_N-1}{2}-1}e^{-\frac{\beta}{2}}\W_{-\frac{\alpha_N-1}{2},\frac{2-\alpha_N}{2}}(\beta)\right)
\end{split}.
\end{equation}
Then, substituting \eqref{eq:e0}$-$\eqref{eq:eN} into \eqref{eq:ae}, the average harvested RF energy $\varepsilon_{\rm avg}$ is obtained. 

To investigate the difference between the average harvested RF energy with BAE and without BAE, we need to give the average harvested RF energy without BAE. As the beam angle differences from the non-associated ETxs in $\Phi_L$ or $\Phi_N$ are inevitable \cite{IBA}, the difference of average harvested RF energy only happens in $\E\{\varepsilon_0\}$. Apparently, if there is no BAE, the probability of $\widetilde \Omega_{x_0}=1$ equals to 1. Therefore, by \eqref{eq:e0}, $\E\{\varepsilon_0\}$ on the condition of   $\widetilde \Omega_{x_0}=1$  is 
\begin{equation}
\E\{\varepsilon_0|\widetilde\Omega_{x_0}=1\}=P_tG_m^2 C_Lr^{-\alpha_L}_0 .
\end{equation}
Then, we define the relative energy loss (REL) of average harvested RF energy in the IBA scenario as 
\begin{equation}\label{eq:delta}
\begin{split}
{\Delta_\varepsilon}=&\frac{\E\{\varepsilon_0|\widetilde\Omega_{x_0}=1\}-\E\{\varepsilon_0\}}{\E\{\varepsilon_0|\widetilde\Omega_{x_0}=1\}}\\=& 1-\frac{1-g^{\frac{1}{2\eta \sigma^2}+1}}{(2\eta \sigma^2+1){\rm erf}^2\left(\frac{\pi }{\sqrt{2\sigma^2}}\right)},
\end{split}
\end{equation} where $\Delta_{\varepsilon}\in [0,1]$. 
If $\Delta_{\varepsilon}=0$, it shows the BAE incurs no energy loss compared to the ideal case, i.e., perfect beam alignment. While, if $\Delta_{\varepsilon}=1$,  it means no energy can be harvested by the typical ERx with BAE. Observe \eqref{eq:delta}, if $\sigma^2\rightarrow 0$,  there is $\Delta_\varepsilon\rightarrow 0$. That is to say  the derived $\varepsilon_{\rm avg}$ can cover the perfect beam alignment case.  

\section{Simulation Results}
In this section, we verify our theoretical results by Monte Carlo simulations. The carrier frequency is  $28$ GHz. Unless otherwise specified, the system parameters are listed in Table \ref{ParaTable}. These values are based on the simulation parameters in \cite{nonlinear, Renzo}. The antenna pattern parameters in our simulations are shown in Table \ref{table:antenna}.  In the figure legends, \textbf{`Theory'} means the theoretical results obtained by our derived analytic expressions and others are the results from the Monte Carlo simulations. For comparison, we also simulated the performance of mmWave WPT system with the flat-top antenna model and the actual beam pattern of ULA model. To let the flat-top model have the same maximum mainlobe gain and 3dB mainlobe beamwidth as the Gaussian antenna model for fair comparison, we define the flat-top antenna model as 
\begin{equation}
G_{\rm F}(\theta)=\begin{cases}
G_m, & |\theta|\leq \theta_{\rm 3dB},\\
G_s, & |\theta_{\rm 3dB}|< |\theta| \leq \pi.
\end{cases}
\end{equation} While, by \cite{sin_antenna},\cite{antenna_book}, the actual beam gain of ULA model can be written as 
\begin{equation}\label{eq:gula}
	G_{\rm U}(\theta)=N_a\left(\frac{\sin^2(\frac{N_a\theta}{2})}{N_a^2\sin^2(\frac{\theta}{2})}\right), 0\leq |\theta|\leq \pi,  
\end{equation} where $N_a$ is the antenna number of ULA and must be an integer. Since it is hard to let the ULA model have the same maximum mainlobe gain and mainlobe beamwidth as the Gaussian antenna model, we herein force both models to achieve the same mainlobe beamwidth $2\theta_0$. According to \cite{antenna_book}, we let $N_a$ be approximately equal to $[ 5.64/\theta_0]_{\mathbb{Z}}$, where $[x]_{\mathbb{Z}}$ is the closest integer to the real number $x$. For $\theta_0=\pi/6$, $\pi/12$, $\pi/24$, therefore, $N_a$ shall be $11$, $22$, $43$, respectively. In addition, by comparing \eqref{eq:ngain} with \eqref{eq:gula}, the roll-off factor of ULA antenna model is slightly different from that of Gaussian model.
\begin{table}[tb]
\caption{Parameters in Simulations}
\label{ParaTable}
\centering
  \begin{tabular}{l l l }
  \toprule[1.5pt]
  \bf{Symbol} & \bf{Definition} & \bf{Default Value}  \\
  \midrule
  $P_t$ &  ETx transmit power & 40 dBm\\
  $\lambda_t$ & Density of ETxs  & $5\times 10^{-4}/$m$^2$\\
  $r_0$ & Distance between Typical ETx-ERx & 50  m\\
  $\kappa$ & Spacing distance/wavelength  ($d/\nu$)  & 0.25\\
  $\alpha_L$ & Path loss exponent of LOS & 2.1 \\
  $\alpha_N$ & Path loss exponent of NLOS & 2.92  \\
  $C_L$ & Path loss intercept of LOS & $10^{-\frac{61.4}{10}}$\\
  $C_N$ & Path loss intercept of NLOS &  $10^{-\frac{72}{10}}$\\
  $M_L$ &  Gamma fading parameter of LOS& 3\\
  $M_N$ &  Gamma fading parameter of NLOS& 2\\
  $\beta$ & Blockage parameter & $0.0071$ \\
  $p_m$ & Maximum harvested power  & 10 mW\\
  $p_a$ & Circuit parameter   &  1500\\
  $p_b$ & Circuit parameter   &  0.0022\\
$K$ & Gamma approximation parameter & $5$  \\
  \bottomrule[1.5pt]
\end{tabular}
\end{table}

\begin{table}
\caption{Parameters of Gaussian Antenna Gain}
\label{table:antenna}
\centering
\begin{spacing}{1.5}
\begin{tabular}{|c|c|c|c|c|}
\hline
$\theta_0$& $\theta_{\rm 3dB}=\frac{\theta_0}{2.6} $& $\eta=\frac{0.3\ln 10}{\theta_{\rm 3dB}^2}$ &  $G_m= \frac{\pi 10^{2.028}}{42.6443\theta_0+\pi}$ \\ 
\hline 
$\pi/24$ & 0.0503& 272.5250 & 38.4103\\ 
\hline 
$\pi/12$ &0.1007 &  68.1313 &23.4227\\ 
\hline 
$\pi/6$ & 0.2014 &17.0328 &    13.1559 \\ 
\hline
\end{tabular} 
\end{spacing}
\end{table}

\begin{figure}[!tb]
\centering
\includegraphics[scale=0.58]{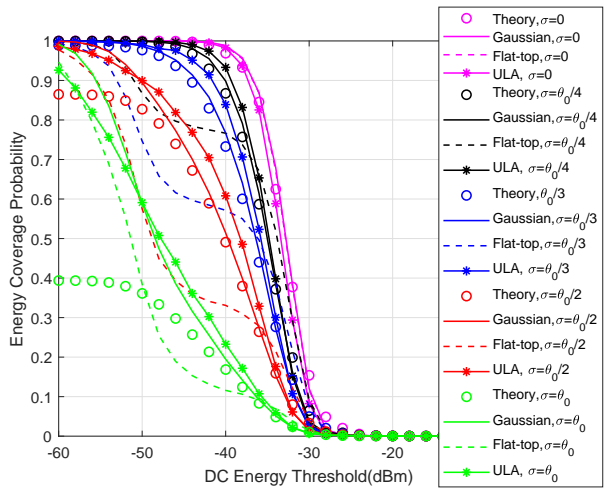}
\caption{Energy coverage probability versus DC energy  threshold, $\theta_0=\frac{\pi}{12}$, $N_a=22$.}
\label{fig:ecp_th}
\end{figure}

In Fig. \ref{fig:ecp_th}, we show the energy coverage performance of  the mmWave  WPT systems with  $\theta_0=\frac{\pi}{12}$ in various IBA scenarios.  First, we can see that in the perfect beam alignment case, i.e., $\sigma=0$, the flat-top and Gaussian models achieve the same energy coverage probabilities and the theoretical curve matches the simulation curve very well. As the ULA model has a different maximum mainlobe gain from the flat-top and Gaussian antenna models, it achieves the slightly different performance when $\sigma=0$.
When $\sigma=\frac{\theta_0}{4}, \frac{\theta_0}{3}$, the theoretical results approach the simulation results of Gaussian antenna model closely. While, for $\sigma=\frac{\theta_0}{2}$ and $ \theta_0$, the gap between the theoretical and simulation curves appears at the low threshold regime and it gets larger when $\sigma$ grows. This phenomenon is consistent with the observations in Fig. 2. The reason is that we ignore the sidelobe gain of the Gaussian BAE model.    
Therefore, it can be concluded  that if $P_0^G(\theta_0)\cdot P_0^G(\theta_0) \approx 1$, the derived analytic expression of the energy coverage probability can accurately evaluate the performance of  the considered mmWave WPT systems. 

On the other hand, for $\sigma\neq 0$, the energy coverage performance of the flat-top model is drastically different from that of the 3GPP Gaussian model we used. It reveals that the flat-top antenna model is not suitable for evaluating the  performance of the mmWave WPT systems in the IBA scenario. Moreover, for $\sigma=\theta_0/4, \theta_0/3, \theta_0/2, \theta_0$, the Gaussian antenna model achieves similar energy coverage probability to the ULA model, especially in small $\sigma$ cases. The performance difference  between Gaussian antenna model and ULA model in the imperfect beam alignment case mainly results from the slight  divergence on antenna gain and roll-off factor. Therefore, the Gaussian antenna model can be regarded as a useful and tractable mathematical tool to analyze the system-level performance of mmWave WPT networks in the presence of BAE, while guaranteeing a certain degree of accuracy. We can also see that the larger $\sigma$ is the lower $P_{ec}$ appears.

\begin{figure}[tb]
\centering
\includegraphics[scale=0.58]{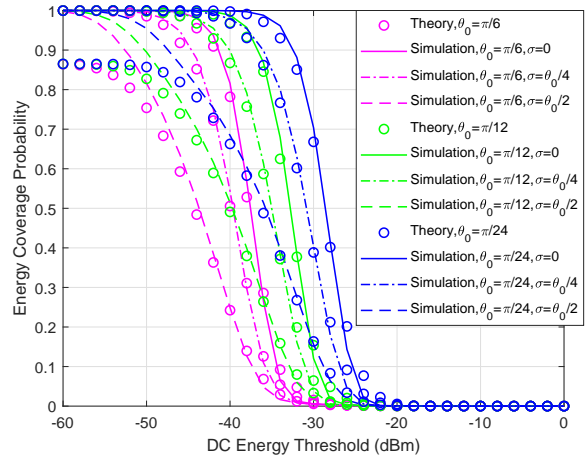}
\caption{Energy coverage probability with various mainlobe beamwidths versus DC energy threshold.}
\label{fig:ecp_width}
\end{figure}
To investigate the effect of mainlobe beam width $\theta_0$, we exhibit the energy coverage performance of the Gaussian antenna model with $\theta_0=\frac{\pi}{24}, \frac{\pi}{12}, \frac{\pi}{6}$ in Fig.  \ref{fig:ecp_width}.   Firstly, for $\sigma=0$ and $ \frac{\theta_0}{4}$, all theoretical results generated by our derived expression match the simulation results closely. So, these curves verify the theoretical results. For $\sigma=\frac{\theta_0}{2}$, the theoretical results in all three $\theta_0$ cases generate  nearly the same gap compared with the corresponding simulation results. 
Secondly, as the threshold $\varepsilon_{\rm th}$ increases, the energy coverage probability decreases in all cases. Comparing  the curves with the same mainlobe width, we can also see that the BAE indeed degrades the performance of the mmWave WPT system. For example, in the scenario of $\theta_0=\frac{\pi}{24}$, the energy coverage probability with $\sigma=\frac{\theta_0}{4}$ reduces from $0.78$ to $0.51$ when $\varepsilon_{\rm th}=-40$ dBm. It means that minimizing BAE is one of the most crucial issues for  mmWave WPT systems.  Besides, using the analytic expression of $P_{ec}$, we can choose the proper mainlobe beamwidth $\theta_0$, if  $P_{ec}$ and $\sigma$ are given. 

\begin{figure}[!tb]
\centering
\includegraphics[scale=0.58]{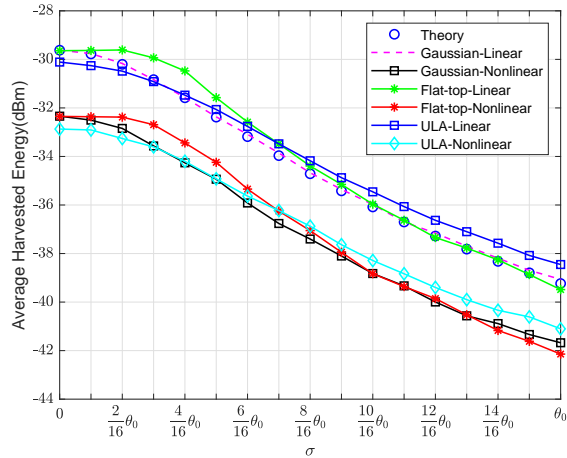}
\caption{Average harvested energy versus BAE standard deviation, $\theta_0=\frac{\pi}{12}, N_a=22,  \lambda_t=10^{-4}/m^2$.}
\label{fig:ae_sigma}
\end{figure}

Fig. \ref{fig:ae_sigma} illustrates the effect of the BAE standard deviation $\sigma$ on the average harvested energy. We consider six mmWave WPT systems, i.e.,  Gaussian antenna model with linear EH, Gaussian antenna model with nonlinear EH, Flat-top antenna model with linear EH, Flat-top antenna model with nonlinear EH, ULA model with linear EH, and ULA model with nonlinear EH. Apparently, as $\sigma$ increases, the average harvested energy decreases. Observing the purple dashed line, the theoretical average energy fits the simulation results very well. It verifies our derived close-form expression of the average harvested RF energy. Furthermore, regardless of linear or nonlinear EH model, the flat-top antenna model has quite different results from the Gaussian antenna model and the ULA model in small $\sigma$ cases, i.e., $\sigma < \frac{10}{16}\theta_0$. When $0<\sigma < \frac{10}{16}\theta_0$, the flat-top antenna model gains more energy than the other two  antenna models. Nevertheless, as $\sigma$ grows, such as $\sigma>\frac{8}{16}\theta_0$, the ULA  model gains more energy than the other two models. Given $\sigma$, the linear EH model always harvests more energy than the nonlinear EH model. This is because we set the RF-DC conversion efficiency of linear EH model as 1. From this point, we can also conclude that the Gaussian antenna model is an  accurate approximation of the ULA model for the small $\sigma$ situations, and the performance gap between both models is acceptable for the sake of analysis tractability. 

\begin{figure}[!tb]
\centering
\includegraphics[scale=0.58]{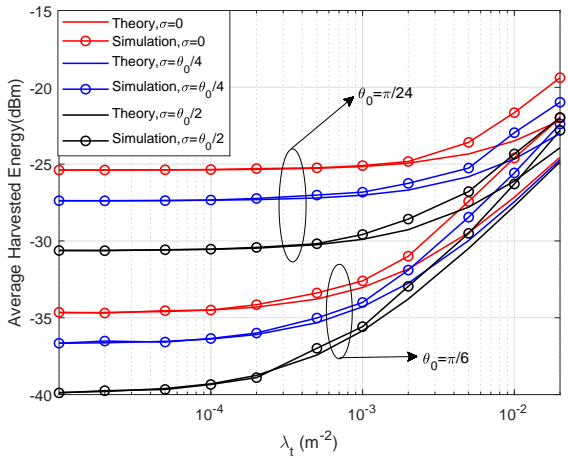}
\caption{Average harvested energy versus density of ETxs.}
\label{fig:ae_lambda}
\end{figure}

Fig. \ref{fig:ae_lambda} illustrates the performance of the average harvested energy with different mainlobe widths. To demonstrate these curves distinctively, we herein take $\theta_0=\frac{\pi}{24}, \frac{\pi}{6}$ as representatives. When $\lambda_t \leq 2\times 10^{-4}$, all  theoretical results match the simulation results exactly.  As $\lambda_t $ increases, the average harvested energy begins to grow. The reason is that the average distance between ETx and ERx gets closer. Additionally, in this case the sidelobe gain plays a more and more significant role in the average harvested energy and the theoretical results based on the approximated PDFs start to be less than the simulation results.  Certainly, enlarging $\lambda_t$ can  compensate the performance degradation caused by BAE. 

\begin{figure}[!tb]
\centering
\includegraphics[scale=0.58]{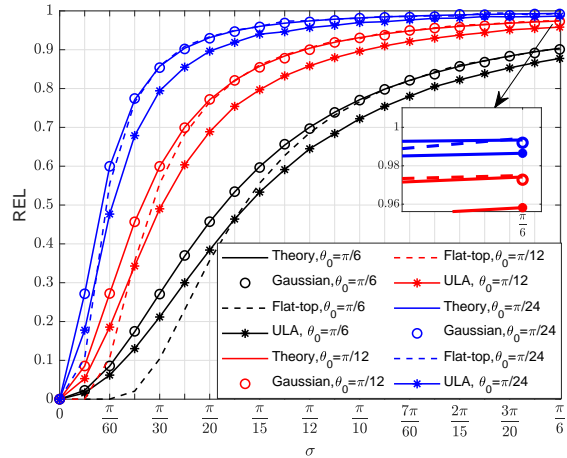}
\caption{Relative energy loss versus BAE standard deviation, $\lambda_t=10^{-4}/m^2$.}
\label{fig:rel}
\end{figure}

In Fig. \ref{fig:rel},  we show the performance of REL versus $\sigma$. First, the theoretical curves match the simulation curves exactly. As we stated earlier, when $\sigma=0$, all RELs equal to zero, which means no energy loss incurred by BAE in those cases.  As $\sigma$ increases, REL increases in all scenarios. With the same $\sigma$, we can see the mmWave WPT system with $\theta_0=\pi/24$ produces the largest REL among all three cases. Therefore, we can deem that the BAE leads to larger performance degradation for mmWave WPT system with stronger directional antenna. Besides, given $\theta_0$, the flat-top model incurs lower REL than the Gaussian model in the small $\sigma$ regime, e.g., for $\theta_0=\frac{\pi}{12}$ and $\sigma <\frac{\pi}{20}$. Differently, the REL of the ULA model approaches that of Gaussian antenna model closely when $\sigma<\frac{\pi}{60}$. Moreover, the REL of ULA model is always less than those of Gaussian and flat-top antenna models. The performance gap mainly results from the difference in the maximum mainlobe gain, the sidelobe gain, and roll-off factor between the ULA model and the other two antenna models.
It is worth mentioning that for $\theta_0=\frac{\pi}{24}$, even when $\sigma=\frac{\pi}{6}$, i.e., $\sigma=4\theta_0$, the REL is still less than 1. This is because the average harvested energy can not be zero no matter how severe the BAE is.

\section{Conclusions}
The impact of imperfect beam alignment on wireless power transfer at millimeter wave frequencies has been investigated in this paper.  The beam alignment error (BAE) from the associated  ETx-ERx transmission is modeled as the truncated Gaussian distribution, while, the BAE from the non-associated ETx-ERx transmission follows the uniform distribution. Then, we derive the probability density functions of the cascaded antenna gains with both mentioned stochastic BAE models and their approximated expressions with more tractability are also provided.  The analytic expression of energy coverage probability has been derived.  Moreover, we also give the closed-form expression of average harvested energy under linear energy harvesting model. 
 Finally, the simulation results verify our theoretical expressions. 
\begin{appendices}
\section{Proof of Lemma 1}
\setcounter{equation}{0}
\renewcommand{\theequation}{A.\arabic{equation}}
Observe \eqref{eq:ngain}, when $|\psi|\leq \theta_0$,  $\widetilde{G}(\psi)$ is a  continuous r.v. with respect to $\psi$. While if  $\theta_0<|\psi|\leq \pi$, $\widetilde{G}(\psi)$ is a  discrete r.v. with  probability mass function (PMF) $\Pr{(\widetilde{G}=g)}=\Pr(\theta_0<|\psi|\leq \pi)$. Therefore, $\widetilde{G}(\psi)$ is the mixed r.v. over $[g,1]$.  

Firstly, we derive the PDF of the continuous component.  Denote the normalized mainlobe gain as $\widetilde{G}_m=e^{-\eta \psi^2 }$.  For an arbitrary  $y \in [g,1]$, the CDF of  $\widetilde{G}_m$ is given by
\begin{equation*}
\begin{split}
&F_{\widetilde{G}_m}(y)=\Pr(e^{-\eta \psi^2 }\leq y, |\psi|\leq \theta_0)=\\
&\Pr\left(\sqrt{\frac{-\ln y}{\eta}}\leq |\psi|\leq \theta_0\right)=F_{|\psi|}(\theta_0)-F_{|\psi|}\left(\sqrt{\frac{-\ln y}{\eta}}\right)
\end{split}
\end{equation*} where $F_{|\psi|}(y)=\int_{-y}^{y}f_{\psi}(y)\d y$ is the CDF of $|\psi|$.  Then, the PDF of $\widetilde{G}_m$ can be written as
\begin{equation}
\begin{split}
f_{\widetilde{G}_m}(y)&=\frac{\d F_{\widetilde{G}_m}(y)}{\d y}=\frac{1}{y\sqrt{-\eta\ln y }}f_{\psi}\left(\sqrt{\frac{-\ln y}{\eta}}\right).
\end{split}
\end{equation}

Denoting the normalized sidelobe gain as $\widetilde{G}_s$, the generalized PDF of $\widetilde{G}_s$ can be expressed as  \cite{random_book},
\begin{equation}
f_{\widetilde{G}_s}(y)=\Pr(\widetilde{G}=g)\delta(y-g)=(1-P_0)\delta(y-g),
\end{equation} in which $P_0=F_{|\psi|}(\theta_0)$. 
So, we have $f_{\widetilde{G}}(y)=f_{\widetilde{G}_m}(y)+f_{\widetilde{G}_s}(y)$.  Then, Lemma 1 is proven. 

\section{Proof Theorem 1}
\setcounter{equation}{0}
\renewcommand{\theequation}{B.\arabic{equation}}

As $\widetilde{G}(\phi_x)$ and $\widetilde{G}(\varphi_x)$ independently follow the PDF $f_{\widetilde{G}}(y)$ in \eqref{eq:g_g}, 
 the PDF of  $\widetilde \Omega_x$ can be given by
\begin{equation}
f_{\widetilde \Omega_x}(\Omega)=\int_g^{1}f_{\widetilde{G}}(y)f_{\widetilde{G}}\left(\frac{\Omega}{y}\right)\frac{1}{y}\d y
\end{equation}
For convenience, in \eqref{eq:g_g} we denote the two terms on the right side of the equal sign as $f_{\widetilde G}^{I}(y)$ and $f_{\widetilde{G}}^{II}(y)$, i.e.,
\[
f_{\widetilde G}^{I}(y)=\frac{1}{\sqrt{2\pi \eta \sigma^2} \erf{\frac{\pi}{\sqrt{2\sigma^2}}}}\frac{y^{\frac{1}{2\eta \sigma^2}-1}}{\sqrt{-\ln y}},~~~y\in [g,1]
\]
\[
f_{\widetilde G}^{II}(y)=(1-P_0^G)\delta(y-g)
\]
Obviously, $f_{\widetilde{\Omega}_x}(\Omega)=\mathcal{F}_{1}(\Omega)+2\mathcal{F}_{2}(\Omega)+\mathcal{F}_{3}(\Omega)$, where 
\begin{equation*}
\begin{split}
&\mathcal{F}_{1}(\Omega)=\int_g^{1}f_{\widetilde G}^I(y)f_{\widetilde G}^{I}\left(\frac{\Omega}{y}\right)\frac{1}{y}\d y,\\
&\mathcal{F}_{2}(\Omega)=\int_g^{1}f_{\widetilde G}^{I}(y)f_{\widetilde G}^{II}\left(\frac{\Omega}{y}\right)\frac{1}{y}\d y,\\
&\mathcal{F}_{3}(\Omega)=\int_g^{1}f_{\widetilde G}^{II}(y)f_{\widetilde G}^{II}\left(\frac{\Omega}{y}\right)\frac{1}{y}\d y.
\end{split}
\end{equation*}

Since the domain of $f_{\widetilde G}^I(y)$ is $g\leq y \leq 1$, there is  $\Omega \leq y \leq {\Omega}/{y}$. Thus, we can obtain $\mathcal{F}_{1}(\Omega)$ in two cases. If $\Omega \in[g^2,g)$,  we have
\begin{equation}\label{eq:f1}
\begin{split}
\mathcal{F}_{1}(\Omega)&=\frac{\Omega^{\frac{1}{2\eta \sigma^2}-1}}{2\pi \eta \sigma^2 \erf{\frac{\pi}{\sqrt{2\sigma^2}}}}\int_{g}^{\frac{\Omega}{g}}\frac{1}{\sqrt{-\ln y}}\frac{1}{\sqrt{-\ln \frac{\Omega}{y}}}\frac{1}{y}\d y\\
&=\frac{\Omega^{\frac{1}{2\eta \sigma^2}-1}}{\pi \eta \sigma^2 \erf{\frac{\pi}{\sqrt{2\sigma^2}}}}\arctan\left(\frac{\ln\Omega-2\ln g }{2\sqrt{\ln\Omega \ln \frac{\Omega}{g}}}\right)
\end{split}.
\end{equation}
If  $\Omega \in[g,1]$, there is 
\begin{equation}
\begin{split}
\mathcal{F}_{1}(\Omega)&=\frac{\Omega^{\frac{1}{2\eta \sigma^2}-1}}{2\eta \sigma^2 \erf{\frac{\pi}{\sqrt{2\sigma^2}}}}
\end{split}.
\end{equation}  
And then we have  
\begin{equation}
\begin{split}
\mathcal{F}_{2}(\Omega)&=\int_g^{1}(1-P_{0}^G)\delta(y-g)f_{G}^{II}\left(\frac{\Omega}{y}\right)\frac{1}{y}\d y\\
&=\frac{(1-P_{0}^G)g^{-\frac{1}{2\eta \sigma^2}}}{\sqrt{2\pi \eta \sigma^2} \erf{\frac{\pi}{\sqrt{2\sigma^2}}}}\frac{\Omega^{\frac{1}{2\eta \sigma^2}-1}}{\sqrt{-\ln 
\frac{\Omega}{g}}}, ~~\Omega\in [g^2,g)
\end{split}
\end{equation}
Furthermore, there is 
\begin{equation}
\begin{split}
\mathcal{F}_{3}(\Omega)&=\left(1-P_{0}^G\right)^2\int_g^1 \delta(y-g)\delta\left(\frac{\Omega}{y}\right)\frac{1}{y}\d y\\
&=(1-P_{0}^G)^2\delta(\Omega-g^2)
\end{split}
\end{equation}
According to the domains of $\mathcal{F}_1(\Omega)$,  $\mathcal{F}_2(\Omega)$ and  $\mathcal{F}_3(\Omega)$, we can achieve the overall  PDF  of $\Omega$ as \eqref{eq:pdf_g}. 
\end{appendices}
.
\section*{Acknowledgment}
The authors would like to thank the anonymous reviewers for their valuable comments and suggestions to improve the expression and quality of this paper. This work has also benefited from suggestions by Dr. Zhengdao Wang.

\ifCLASSOPTIONcaptionsoff
  \newpage
\fi

\bibliographystyle{IEEEtran}
\bibliography{ref}

\end{document}